\documentclass[reprint,aps,prx,longbibliography,superscriptaddress,twocolumn,10pt]{revtex4-2}

\usepackage{graphicx}
\usepackage{amssymb}
\usepackage{siunitx}
\usepackage{physics}

\makeatletter
\def\maketitle{
\@author@finish
\title@column\titleblock@produce
\suppressfloats[t]}
\makeatother

\begin{document}

\title{Andreev-enhanced conductance quantization and gate-tunable induced superconducting gap in germanium}

\author{Elyjah Kiyooka}
\email{elyjah.kiyooka@polytechnique.edu}
\altaffiliation{Present address:  QCMX Lab, Laboratoire de Physique de la Matière condensée, CNRS, École polytechnique, Institut Polytechnique de Paris, 91120 Palaiseau, France.}
\affiliation{Univ. Grenoble Alpes, CEA, Grenoble INP, IRIG, PHELIQS, 38000 Grenoble, France}
\author{Chotivut Tangchingchai}
\affiliation{Univ. Grenoble Alpes, CEA, Grenoble INP, IRIG, PHELIQS, 38000 Grenoble, France}
\author{Gonzalo Troncoso Fernandez-Bada}
\affiliation{Univ. Grenoble Alpes, CEA, Grenoble INP, IRIG, PHELIQS, 38000 Grenoble, France}
\author{Boris Brun-Barriere}
\affiliation{Univ. Grenoble Alpes, CEA, Grenoble INP, IRIG, PHELIQS, 38000 Grenoble, France}
\author{Simon Zihlmann}
\affiliation{Univ. Grenoble Alpes, CEA, Grenoble INP, IRIG, PHELIQS, 38000 Grenoble, France}
\author{Romain Maurand}
\affiliation{Univ. Grenoble Alpes, CEA, Grenoble INP, IRIG, PHELIQS, 38000 Grenoble, France}
\author{Francois Lefloch}
\affiliation{Univ. Grenoble Alpes, CEA, Grenoble INP, IRIG, PHELIQS, 38000 Grenoble, France}
\author{Vivien Schmitt}
\affiliation{Univ. Grenoble Alpes, CEA, Grenoble INP, IRIG, PHELIQS, 38000 Grenoble, France}
\author{Jean-Michel Hartmann}
\affiliation{Univ. Grenoble Alpes, CEA, LETI, 38000 Grenoble, France}
\author{Manuel Houzet}
\affiliation{Univ. Grenoble Alpes, CEA, Grenoble INP, IRIG, PHELIQS, 38000 Grenoble, France}
\author{Silvano De Franceschi}
\email{silvano.defranceschi@cea.fr}
\affiliation{Univ. Grenoble Alpes, CEA, Grenoble INP, IRIG, PHELIQS, 38000 Grenoble, France}

\date{\today}

\begin{abstract}
Ge/SiGe quantum well heterostructures confining a high-mobility two-dimensional hole gas (2DHG) have emerged as a compelling platform for hybrid superconductor(S)-semiconductor(Sm) quantum devices. Here, we investigate the low-temperature transport properties of split-gate quantum point contacts (QPC) defined in one such heterostructure and positioned at different distances from an aluminum superconducting contact. We observe ballistic one-dimensional transport evidenced by conductance quantization with at least four clearly visible plateaus.  Andreev reflection at the S/Sm interface induces a 40\% enhancement of the conductance steps relative to the normal-state conductance staircase measured under a 100-mT out-of-plane magnetic field. This result is in excellent agreement with the theoretical expectation for an interface transparency of 0.88. By operating the QPCs in the tunneling regime, we probe the local density of states of the proximitized 2DHG. We report direct experimental evidence of an induced superconducting gap, demonstrating that its magnitude can be tuned by a gate voltage acting  on the carrier density in the 2DHG.     
\end{abstract}

\maketitle

\section{Introduction}

% General field summary
Superconductor/semiconductor (S/Sm) hybrid systems provide a versatile platform for exploring novel quantum phenomena and applications. Their unique functionalities arise from the interplay between electron pairing correlations \cite{Schapers2001}, introduced by the superconductor \cite{Tinkham1996}, and electrically tunable single-particle properties, such as quantum confinement and spin-orbit coupling, offered by the semiconductor \cite{Datta1995}. Over the past two decades, interest in S/Sm hybrid devices has largely increased, primarily driven by their potential applications in solid-state quantum computing. Notable examples include gate-tunable transmons \cite{Casparis2018,Sagi2024,Kiyooka2025}, Andreev qubits \cite{Tosi2019,Hays2021,Pita-Vidal2023}, and significant advances towards topological qubits \cite{Mourik2012,Zatelli2024,TenHaaf2025}.

% Motivate the material
Several proof-of-concept implementations have been demonstrated across various material systems, with a predominant focus on InAs- \cite{Kjaergaard2017,Junger2019}, InSb- \cite{Gul2017,Ke2019}, and InSbAs-based \cite{Wang2024} semiconductor nanostructures and a raising interest for Ge/SiGe heterostructures \cite{Vigneau2019,Tosato2023,Valentini2024}. However, basic aspects such as the spatial distribution and gate tunability of the induced superconducting gap remain largely unexplored. Moreover, certain applications, including topological qubits based on Majorana zero modes or parity-protected superconducting qubits, require ballistic transport in the semiconductor channel \cite{Lutchyn2018, Ahn2021}, a condition that has been hard to meet in small band-gap III-V semiconductors, where carrier mobility remains a clear limitation \cite{Shabani2016}, especially in the one-dimensional limit \cite{Kjaergaard2016}. 

% Experiment summary
In this work, we address these challenges by means of purposely designed S/Sm hybrid devices fabricated from a Ge/SiGe heterostructure embedding a two-dimensional hole gas (2DHG). The devices consist of split-gate quantum point contacts (QPCs) located in the proximity of a superconducting aluminum contact. Owing to the high mobility and low percolation density of the 2DHG, we observe clear evidence of ballistic one-dimensional transport leading to conductance quantization. We find that conductance steps are enhanced by Andreev reflection at the S/Sm interface, in agreement with the theoretical expectation for an interference transmission close to unity. Then, by operating the QPC in a low-conductance, tunneling regime, we perform a spectroscopy of the local density of states (LDOS) in the proximitized quantum-well region. We report direct evidence of an induced superconducting gap and present a study of its spatial and gate-voltage dependence, providing new insight on the superconducting proximity effect.  

%\section{Results/Discussion}

% Brief fabrication section  - Hall data
We fabricate devices from a Ge/SiGe heterostructure that consists of a \SI{16}{\nano\meter} thick strained Ge quantum well between relaxed Si$_{.21}$Ge$_{.79}$ barrier layers with an upper barrier thickness of \SI{22}{\nano\meter}. Additional growth details were reported in an earlier publication \cite{Hartmann2023}. The superconducting contacts are defined through a calibrated dry etching of the upper barrier layer followed by an e-beam evaporation of a \SI{50}{\nano\meter} thick Al layer on top of the uncovered Ge quantum well \cite{Chotivut2024} and a lift-off process. Two overlapping gate layers are defined with intervening steps of Al$_{2}$O$_{3}$ atomic-layer deposition. More details on the fabrication procedure can be found in Appendix A. In similarly fabricated Hall bar devices, we measure the gate dependence of the hole carrier density ($\rho_{\mathrm{2D}}$), and mobility ($\mu$) (see Fig. S1). These values allow us to estimate the inelastic mean free path ($l_{\mathrm{mfp}}$) which extends up to 1.8 \SI{}{\micro\meter} in the high density regime ($\rho_{\mathrm{2D}} \sim 7\times 10^{11}$\SI{}{\centi\meter}$^{-2}$).

\begin{figure}[ht!]
	\centering
	 \includegraphics[width=\linewidth]{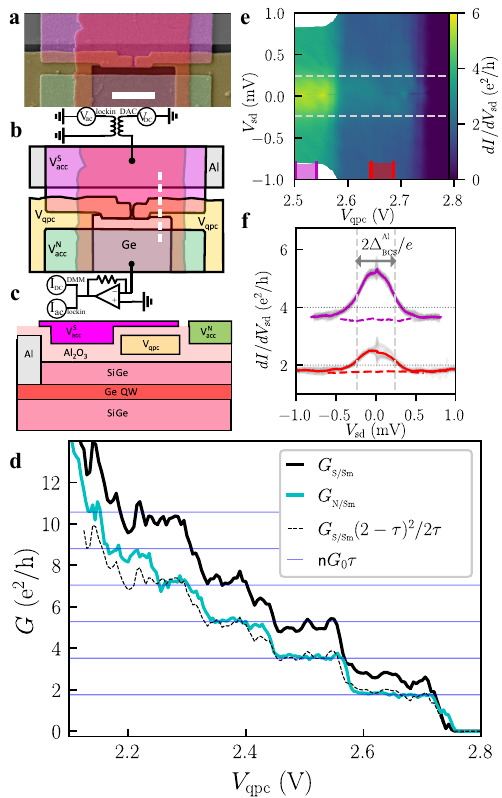}
\caption{
%Conductance quantization in a quantum-point contact (QPC) facing an Al superconducting electrode. 
(a) False-colored scanning-electron micrograph (SEM) of device D1 (scale bar: \SI{1}{\micro\meter}). (b) Top-view device schematic with simplified measurement circuitry. (c) Cross-sectional schematic taken along the dashed line in (b). (d) Linear conductance, $G$, vs QPC gate voltage, $V_{\mathrm{qpc}}$. Cyan solid line: normal-state conductance, $G_{\mathrm{N/Sm}}(V_{\mathrm{qpc}})$, measured at a perpendicular magnetic field  $B_{\perp}$=\SI{100}{\milli\tesla}. Black solid line: Andreev-enhanced conductance, $G_{\mathrm{S/Sm}}(V_{\mathrm{qpc}})$, measured at zero magnetic field. Black dashed line: measured Andreev-enhanced conductance scaled by the theoretically expected ratio $G_{\mathrm{S/Sm}}/G_{\mathrm{N/Sm}} =  2\tau/(2 - \tau)^2$ for a fitted transmission coefficient $\tau = 0.88$. $G_{\mathrm{N/Sm}}$ and the scaled $G_{\mathrm{S/Sm}}$ exhibit overlapping  plateaus well aligned with a set of horizontal thin lines positioned at integer multiples of $\tau G_0$, again with $\tau = 0.88$. (e) Color plot of $dI/dV_{\mathrm{sd}}(V_{\mathrm{qpc}}, V_{\mathrm{sd}})$ with dashed lines indicating where the bias voltage matches the Al superconducting gap.  (f)  Solid lines: $dI/dV_{\mathrm{sd}}(V_{\mathrm{sd}})$ obtained from (e) by averaging line-cut traces from the central regions of the first and second plateau indicated by colored bands. The ensemble of line-cuts used for this averaging are overlayed as light grey lines. We also plot as dashed traces the line-cut averages calculated from similar  $dI/dV_{\mathrm{sd}}(V_{\mathrm{qpc}}, V_{\mathrm{sd}})$ measurements in the absence of superconductivity at $B_{\perp}$=\SI{100}{\milli\tesla}.}  
\end{figure}

% Summary of Device 1 - Fig. 1 (a-c)
The first device, which we refer to as device D1, is displayed in Fig. 1(a) as a top-view, false-colored scanning electron micrograph showing the Ge mesa (red), aluminum contact (gray), QPC gates (yellow), and two accumulation gates on the upper (magenta) and lower side (cyan) of the QPC. For further clarity, we provide in Fig. 1(b) a top-view schematic and in Fig. 1(c) a cross-sectional schematic. The QPC split gates are designed to be \SI{200}{\nano\meter} long with an \SI{80}{\nano\meter} gap between them and a distance of $d_{\mathrm{D1}} \sim$ \SI{300}{\nano\meter} from the superconducting contact. The two accumulation gates control the carrier density in the underlying 2DHG through the applied voltages  $V^{\mathrm{N}}_{\mathrm{acc}}$ and $V^{\mathrm{S}}_{\mathrm{acc}}$. Unless otherwise stated, these voltages are adjusted to achieve full carrier accumulation placing the system in the ballistic regime $l_{\mathrm{mfp}} > d_{\mathrm{D1}}$. Low-temperature transport measurements are performed in a dry dilution refrigerator with a base temperature of \SI{7}{\milli\kelvin}.  

\section{Andreev-enhanced conductance}

Fig. 1(d) shows two-terminal measurements of the linear conductance, $G$, as a function of the voltage, $V_{\mathrm{qpc}}$, applied simultaneously to the two QPC split  gates. The data have been plotted after subtracting the contribution of a series resistance measured when all gates are fully accumulated underneath (see Appendix C, and D). The cyan solid line is a measurement taken in the presence of a magnetic field, $\mathrm{B}_{\perp} = $ \SI{100}{\milli\tesla}, perpendicular to the 2DHG and large enough to suppress superconductivity in the Al contact. This normal-type conductance exhibits clear steps close to integer multiples of the conductance quantum, $G_{0} = 2e^{2}/h$, with $e$ the electron charge and $h$ the Planck constant. Such a conductance staircase is the characteristic signature of ballistic one-dimensional transport \cite{VanWees1988,Gao2024,Hudson2026}. 
The black solid line is a measurement at zero magnetic field, namely in the presence of superconductivity.  Remarkably, conductance quantization is preserved, but the step height increases to $\sim1.25\times G_{0}$. This conductance step enhancement arises from Andreev reflection at the S/Sm interface, as theoretically predicted by Beenakker \cite{Beenakker1992} and, so far, rarely observed in experiments \cite{Kjaergaard2016,Irie2016,Gao2025}.

% Enhancement \tau extraction
Following Ref. \cite{Beenakker1992}, the conductance of a QPC connecting a normal and a superconducting contact can be generically expressed as $G_{\mathrm{S/Sm}} = \sum^{n}_{i=1} G_{0} (2\tau_i^2/(2-\tau_i)^{2}) $, where $\tau_i$ is the transmission coefficient of the $i$-th conduction mode, and $n$ is the total number of conducting modes. When superconductivity is suppressed, the QPC normal-type  conductance is given by $G_{\mathrm{N/Sm}} = \sum^{n}_{i=1} G_{0} \tau_i$. Assuming that all the modes have the same transmission coefficient, $\tau$, the above expressions simplify to $G_{\mathrm{S/Sm}} = n G_{0} (2\tau^2/(2-\tau)^{2}) $ and $G_{\mathrm{N/Sm}} = n \tau G_{0} $, respectively. To verify this, we fit the plateaus of the measured normal-type conductance to  $G_{\mathrm{N/Sm}} = n \tau G_{0} $,  and the plateaus of the measured Andreev-enhanced conductance to  $G_{\mathrm{S/Sm}} = n (2\tau^2/(2-\tau)^{2}) G_{0}$, each time using $\tau$ as the only fitting parameter. For both cases, the best fit over the first four plateaus is obtained for $\tau = 0.88$ (see Fig. S3). Using this value of $\tau$, we plot a set of horizontal lines at integer multiples of $\tau G_0$ in Fig. 1(d). The normal-type conductance exhibits up to six plateau-like structures overlapping with these horizontal lines. Furthermore, for $\tau = 0.88$ we expect $G_{\mathrm{S/Sm}}/G_{\mathrm{N/Sm}} = (2\tau/(2-\tau)^{2}) = 1.4$. When divided by this factor, the Andreev-enhanced conductance overlaps well with the normal conductance up to the fourth plateau. The results of Fig. 1(d) offer a clear experimental demonstration of Beenakker’s formula for the Landauer conductance quantization in a QPC close to a S/Sm interface.  The observed equal transmission probability of the one-dimensional channels suggests a common origin of the observed back scattering most likely associated with the high, but finite transparency of the S/Sm interface.

% Conductance enhancement with bias - Fig. 1 (e)
We now consider the bias-voltage dependence of the QPC differential conductance, $dI/dV_{\mathrm{sd}}$.  Figure 1(e) shows a color-scale plot of $dI/dV_{\mathrm{sd}}(V_{\mathrm{qpc}},V_{\mathrm{sd}})$ in the regime from $dI/dV_{\mathrm{sd}} \ll G_{0}$ to the second conductance plateau. We observe a clear enhancement of the conductance around zero bias in a range roughly set by the Al superconducting gap, which we have highlighted by the horizontal dashed lines at $V_{\mathrm{sd}} =\pm  \Delta^{\mathrm{Al}}_{_\mathrm{BCS}}/e  = \pm$\SI{240}{\micro\V}. This value for the Al gap is estimated from the critical temperature $T_{\mathrm{C}} = $ \SI{1.58}{\kelvin} of an on-chip Al strip-line using the Bardeen-Cooper-Schrieffer (BCS) relation $\Delta^{\mathrm{Al}}_{_\mathrm{BCS}} = 1.76 k_{\mathrm{B}}T_{\mathrm{C}}$ \cite{Tinkham1996}.

% Confirmation of \tau with other dataset - Fig. 1 (f)
Figure 1(f) displays line-cut averages from the middle regions of the first and second plateau as solid red and magneta lines, respectively. The ensemble of individual line-cut traces used in the averaging are overlaid as faint gray lines. Their resulting envelopes illustrate the amplitude of mesoscopic modulations largely removed by the averaging. For both plateaus, $dI/dV_{\mathrm{sd}}(V_{\mathrm{sd}})$ has a maximum at zero bias and gradually decreases with $\abs{V_{\mathrm{sd}}}$. For $\abs{V_{\mathrm{sd}}}$ well above $\Delta^{\mathrm{Al}}_{_\mathrm{BCS}}$, $dI/dV_{\mathrm{sd}}$ levels off at a value close to the normal-state differential conductance measured at $\mathrm{B}_{\perp} = $\SI{100}{\milli\tesla} in the same gate-voltage windows (shown as dashed red and magneta traces). Noteworthy, when divided by a factor 2, the $dI/dV_{\mathrm{sd}}(V_{\mathrm{sd}})$ traces from the second plateau fall approximately on top of the ones from the first plateau (Fig. S4). This shows that the proportionality relation between Andreev-enhanced and normal conductance revealed by Fig. 1(d) in the linear regime holds all across the high-bias, nonlinear regime. 

% discussion part 1
We remark that the Andreev-enhanced conductance traces in Fig. 1(f) cannot be reproduced by a one-dimensional Blonder-Tinkham-Klapwijk (BTK) model \cite{Irie2016,Jakob2000}. In fact, in contrast to our experimental finding, the BTK conductance for an interface transparency $\tau=0.88$ should exhibit an enhanced-conductance relative minimum at $V_{\mathrm{sd}} = 0$\SI{}{\volt} and symmetric peaks at $V_{\mathrm{sd}} = \pm  \Delta^{\mathrm{Al}}_{_\mathrm{BCS}}/e$. We suggest that the observed line-shape of the non-linear conductance is instead due to the presence of an induced superconducting gap, $\Delta^*$, in the proximitized 2DHG, the existence of which will be demonstrated in the second part of this Article. Since, as we shall see, $\Delta^*  \approx $ \SI{73}{\micro\eV}, the conductance peaks associated to the edges of the induced gap are likely to be merged into a single zero-bias peak due to their energy broadening. 

\begin{figure}[ht!]
	\centering
	\includegraphics[width=\linewidth]{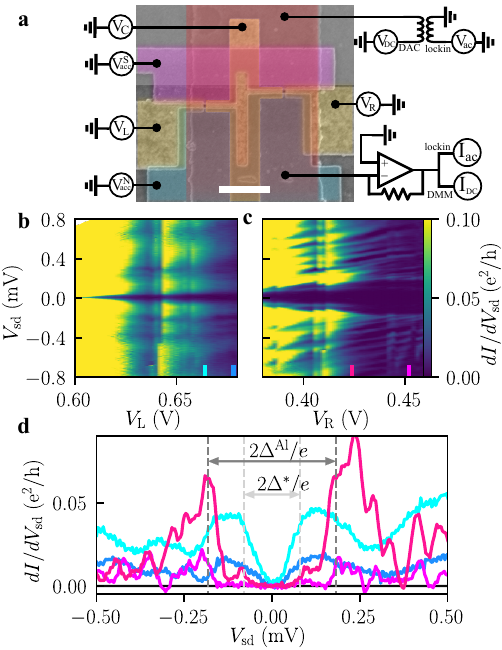}
\caption{Tunnel spectroscopy of the induced superconducting gap in device D2, hosting QPCs at different distances from a same superconducting contact. (a) False-colored SEM image of D2 (scale bar: \SI{1}{\micro\meter}) and simplified measurement circuitry. The image shows two QPCs,  labeled as left and right QPC, respectively at distances  $d_{\mathrm{L}} \sim$ \SI{300}{\nano\meter} and $d_{\mathrm{R}} \sim$ \SI{0}{\nano\meter} from the upper Al contact. (b)  and  (c) show color plots of the differential conductance, $dI/dV_{\mathrm{sd}}$, for the left and right QPC, respectively.  The $dI/dV_{\mathrm{sd}}$ of the left (right) QPC is individually measured as a function of $V_{\mathrm{sd}}$ and the respective QPC gate voltage $V_{\mathrm{L}}$ ($V_{\mathrm{R}}$), while the right (left) QPC is pinched off.  This is achieved with $V_{\mathrm{C}} = $ \SI{0.15}{\volt} ($V_{\mathrm{C}} = $ \SI{0.35}{\volt}) and $V_{\mathrm{R}} = $ \SI{1}{\volt} ($V_{\mathrm{L}} = $ \SI{1}{\volt}). (d) Representative line-cuts from (b) and (c).}
\end{figure}

\section{Induced superconducting gap}

% Introduce part 2
By operating a QPC close to its pinch-off we can use it as a point-like tunnel probe to measure the LDOS in the proximitized region of the 2DHG. This way, we intend to identify clear experimental signatures of a proximity-induced energy gap $\Delta^*$ and measure its dependence on the spatial extent and carrier density of the proximitized region.

% Summary device 2
To this aim, we fabricate device D2 consisting of two superconducting contacts and two nearby pairs of QPCs. In each pair, the QPCs are positioned at different distances from the edge of the nearby superconducting contact. A close-up top view of device D2 is shown in Fig. 2(a), together with a simplified measurement setup. It displays the upper superconducting contact and the two facing QPCs, hereafter referred to as left (L) and right (R) QPC. The QPCs share a common central gate (C), which extends for more than \SI{20}{\micro\meter} down to the bottom superconducting contact, where a qualitatively symmetric gate structure offers the possibility to form two additional QPCs  (a full device image including the bottom contact is presented in Fig. S5). 

% threshold shift part 2
As for device D1, the hole carrier densities on the upper and lower sides of the QPCs can be independently tuned by means of a second layer of top gates. We initially bias these gates to induce strong carrier accumulation all across the gated Ge quantum well. More specifically, we set $V^{\mathrm{S}}_{\mathrm{acc}} = V^{\mathrm{N}}_{\mathrm{acc}} = -2$ V, where $V^{\mathrm{S}}_{\mathrm{acc}}$ and $ V^{\mathrm{N}}_{\mathrm{acc}}$ are the voltages applied to the top gates covering the regions above and below the QPCs in Fig. 2(a), respectively. We notice that the threshold gate voltage was positive for D1, and negative for D2. Based on earlier observations \cite{Troncoso2023,Sangwan2025}, we believe this large negative shift is the effect of an oxygen plasma treatment introduced in the fabrication process of D2 prior to the deposition of the gate dielectric (this treatment was not applied for D1). This change is clearly reflected in the gate-voltage dependence of $\rho_{\mathrm{2D}}$ and $\mu$ from Hall-bar measurements (see Fig. S1).

% measurement specifics
Applying a sufficiently positive voltage $V_{\mathrm{C}}$ to the central gate depletes the 2DHG beneath, separating the 2DHG into two parallel channels that run vertically from the top to the bottom contact. As a result, the left QPC can be individually measured after pinching off the right one, and vice-versa. This device geometry enables tunneling spectroscopy of the LDOS at different distances from the same S/Sm interface. In particular, the left and right QPCs in Fig. 2(a) lie at distances $d_{\mathrm{L}} \approx $ \SI{300}{\nano\meter} and $d_{\mathrm{R}} \approx $ \SI{0}{\nano\meter} from the top superconducting contact, respectively. (These distances are measured between the lower edge of the superconducting contact and the upper edges of the QPC split gates.) Under the reasonable assumption of a uniform transparency along the same S/Sm interface, this configuration ensures that differences in the tunneling spectroscopy measurements can be ascribed to the distance dependence of the proximity effect, and not to device-to-device variations that can arise when comparing QPC–superconductor separations across distinct samples.

% measurement G
Figure 2(b) (2(c)) shows a $dI/dV_{\mathrm{sd}}$ measurement of the left (right) QPC as a function of the voltage applied to the corresponding QPC gate, $V_{\mathrm{L}}$ ($V_{\mathrm{R}}$), and $V_{\mathrm{sd}}$. The contribution from the series resistance - determined from the measurements reported in Fig. S6 - has been subtracted even though it is less relevant in the low-conductance regime of Figs. 2(b) and 2(c). Close to pinch off (right-hand side of Figs. 2(b) and 2(c)), both data sets present a strong $dI/dV_{\mathrm{sd}}$ reduction around zero bias voltage, denoting a gapped LDOS. In the following, assuming the measured $dI/dV_{\mathrm{sd}}$ to be proportional to the LDOS, we shall extract the gap size by fitting the $dI/dV_{\mathrm{sd}}(V_{\mathrm{sd}})$ characteristics to an artificially broadened BCS DOS, using the superconducting gap and the gap-edge broadening as fitting parameters (see Appendix I and Fig. S7). For bias voltages above the gap, $dI/dV_{\mathrm{sd}}(V_{\mathrm{sd}})$ exhibits rather pronounced oscillations that we attribute to Fabry-Perot-type resonances arising from scattering in the QPC region \cite{Irie2016}. To limit the impact of this undesirable mesoscopic effect, we shall systematically limit the fitting $V_{\mathrm{sd}}$ range to the immediate vicinity of the gap.   

\begin{figure}[ht!]
	\centering
	 \includegraphics[width=\linewidth]{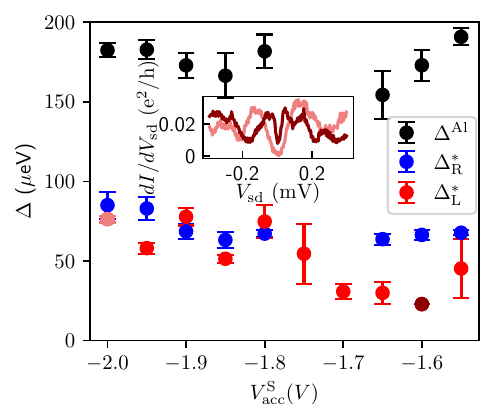}
\caption{Parent and induced superconducting gaps (generically, $\Delta$) extracted from  tunnel spectroscopy of the local density of states in device D2 as a function of the accumulation gate voltage, $V_{\mathrm{acc}}^{\mathrm{S}}$. The induced gap $\Delta^{*}_{\mathrm{R}}$ and the parent gap $\Delta^{\mathrm{Al}}$ are measured using the right QPC, while the induced gap $\Delta^{*}_{\mathrm{L}}$ is measured using the left QPC. All data points are obtained by fitting the $dI/dV_{\mathrm{sd}}$ tunneling characteristics to a broadened Bardeen-Cooper-Schrieffer density of states. Inset: two representative $dI/dV_{\mathrm{sd}}$ traces measured with the left QPC at $V_{\mathrm{acc}}^{\mathrm{S}} = -2$  and \SI{-1.6}{\volt}, corresponding to strong and weak hole accumulation, respectively.}
\end{figure}

% comments on gap structure
In the case of the right QPC, lying closer to the superconducting contact, the gap appears qualitatively wider and presents a sub-gap structure. The outer gap edges can be plausibly associated to the onset of tunneling into (or out of) the quasi-particle states of the Al contact. Their $V_{\mathrm{sd}}$ position is determined by the Al superconducting gap, $\Delta^{\mathrm{Al}}$, at the interface with the Ge quantum well. Our fit to a broadened BCS LDOS yields $\Delta^{\mathrm{Al}} \approx $ \SI{180}{\micro\eV}, which is somewhat smaller than the gap derived from the $T_c$ of the Al strip-line. (We remark that, for high interface transparencies, the gap of the superconductor can significantly shrink near the interface due to inverse proximity effect \cite{Son1988}.) The inner ridges in Fig. 2(c) mark the onset of quasi-particle tunneling above the minigap induced in the quantum-well region below the superconducting contact. Their $V_{\mathrm{sd}}$ position corresponds to a minigap $\Delta^{*}_{\mathrm{R}}\approx $ \SI{80}{\micro\eV}. In the present case of a thin semiconductor quantum well, the minigap can amount to a significant fraction of the Al superconducting gap with a ratio $\Delta^{*}/\Delta^{\mathrm{Al}}_{\mathrm{BCS}}$ determined by the interface transparency and the Fermi-velocity mismatch \cite{Reeg2016,Volkov1995}. 

% introduce thouless energy - motivate length dependence
When the proximitized quantum well is not entirely covered by the superconductor, another energy scale comes into play. This energy, known as the Thouless energy, $E_{\mathrm{Th}}$, is inversely proportional to the characteristic time quasiparticles take to move across the uncovered region of the quantum well. As this time increases, the Thouless energy becomes the smallest energy scale and a suppression of the induced gap is expected at the edge of the two-dimensional semiconductor system, where $ \Delta^{*} \approx E_{\mathrm{Th}}$ \cite{Gupta2004}. In the ballistic limit, $E_{\mathrm{Th}} = h v_{\mathrm{F}}/2 \pi d$, where $v_F$ the Fermi velocity and $d$ the lateral size of the uncovered quantum well region. In the situation probed by the left QPC in Fig. 2(a), we have $d = d_L \approx $ \SI{300}{\nano\meter}. The Fermi velocity can be derived from the estimated carrier density, $\rho_{\mathrm{2D}}$,  using the relation $v_{\mathrm{F}}= (h/m^*) \sqrt{\rho_{\mathrm{2D}}/2\pi}$, where $m^*$ is the hole effective mass, which we expect to be approximately ten percent of the bare electron mass \cite{Lodari2019,KiyookaPhD2025}. In the limit of strong hole accumulation for D2, imposed by $V^{\mathrm{S}}_{\mathrm{acc}}=-2$ V, we estimate $\rho_{\mathrm{2D}}\sim 4\times10^{11}$\SI{}{\centi\meter}$^{-2}$, which gives $v_{\mathrm{F}} \sim 1.8 \times 10^5$ \SI{}{\meter\per\second}, and hence $E_{\mathrm{Th}} \sim $ \SI{0.4}{\milli\eV}. Such a large Thouless energy should not be limiting the minigap as confirmed by the tunnel spectroscopy data of the left QPC. Indeed the data of Fig. 2(b) reveal the presence of a zero-bias dip corresponding to a minigap $\Delta^{*}_{\mathrm{L}} \approx $ \SI{80}{\micro\eV}, hence identical to $\Delta^{*}_{\mathrm{R}}$. For the first device of Fig. 1, having similar $d_{\mathrm{D1}} \approx$ \SI{300}{\nano\meter}, we find a comparable value of the induced gap $ \approx $ \SI{73}{\micro\eV}, as shown in Fig. S8. Figure 2(d) offers a direct comparison between a few representative line cuts from Figs. 2(b) and 2(c).  We notice that tunnel spectroscopy through the left QPC shows no signatures of direct quasiparticle tunneling into the Al contact, which is likely a consequence of the  large distance $d_L$. 

% figure 3 results 
Since $E_{\mathrm{Th}} \propto \sqrt{\rho_{\mathrm{2D}}}$, increasing $V^{\mathrm{S}}_{\mathrm{acc}}$ should reduce $E_{\mathrm{Th}}$. As $E_{\mathrm{Th}}$ falls below \SI{80}{\micro\eV}, its further gate-induced reduction should manifest as a progressive suppression of the minigap $\Delta^{*}_{\mathrm{L}}$. This effect finds experimental confirmation in the results of Fig. 3, where the induced gap $\Delta^{*}_{\mathrm{L}}$ obtained from tunnel spectroscopy is plotted as a function of $V^{\mathrm{S}}_{\mathrm{acc}}$ (red data points). Despite large fluctuations, most likely due to mesoscopic interference effects, $\Delta^{*}_{\mathrm{L}}$ exhibits a clear tendency to decrease with $V^{\mathrm{S}}_{\mathrm{acc}}$, down to a lowest measurable value of $\sim$ \SI{20}{\micro\eV}.  The figure inset shows representative differential-conductance traces for $V^{\mathrm{S}}_{\mathrm{acc}} = $ \SI{-2}{\volt} and $V^{\mathrm{S}}_{\mathrm{acc}} = $ \SI{-1.6}{\volt}, corresponding to the largest and smallest values of $\Delta^{*}_{\mathrm{L}}$, respectively. Figure 3 also shows the gate-voltage dependence of $\Delta^{\mathrm{Al}}$ and $\Delta^{*}_{\mathrm{R}}$ as obtained from tunnel-spectroscopy measurements through the right QPC in the same $V^{\mathrm{S}}_{\mathrm{acc}}$ range (black and blue data points, respectively). The measurement datasets of the left and right QPC for every $V^{\mathrm{S}}_{\mathrm{acc}}$ value are shown in Appendix K. Both $\Delta^{\mathrm{Al}}$ and $\Delta^{*}_{\mathrm{R}}$ remain approximately constant, denoting a negligible field-effect in the quantum-well region under the superconducting contact. Consistent results are found for another QPC device, D3, as shown in Appendix L.

% Discussion of figure 2/3 
The studied gate-voltage dependence of the proximity-induced superconducting gap in a semiconductor channel is the core operating principle of the Josephson field-effect transistor and other related quantum devices, such as gatemons \cite{Valentini2024,Kiyooka2025}, bolometers \cite{Kokkoniemi2020}, and a variety of devices for superconducting cryoelectronics \cite{Buccheri2025,Paghi2025}.  Yet direct experimental studies of the electrostatic tunability of an induced superconducting gap are rather scarce in the literature \cite{Junger2019, DeMoor2018, VanLoo2023}. Moreover, a common challenge lies in the difficulty to discriminate the field effect on the carrier density from that on the S/Sm interface barrier \cite{VanLoo2023}. In our experiment, the device design and the observed gate-voltage independence of $\Delta^{\mathrm{Al}}$ and $\Delta^{*}_{\mathrm{R}}$ in Fig. 3 allow us to rule out a gate effect on the S/Sm interface. 

\section{Conclusions}

% Summary
In summary, we have used gate-defined QPCs to investigate the superconducting proximity effect in a high-mobility Ge quantum-well contacted by superconducting aluminum electrodes. Owing to the ballistic character of the confined 2DHG, we observed clear conductance quantization in the few-mode QPC regime. The first four conductance steps showed an equal enhancement by Andreev-reflection at the S/Sm interface, consistent with a mode-independent interface transmission coefficient of 0.88. Our findings underscore the potential of Ge/SiGe heterostructures to realize proximitized one-dimensional ballistic systems,  a key milestone for achieving topological superconductivity and enabling a variety of fundamental experiments and quantum functionalities. By operating split-gate QPCs in the tunneling regime, we could probe the superconducting gap $\Delta^{*}$ induced in the LDOS of the proximitized 2DHG. Despite undesirable mesoscopic fluctuations in the QPC conductance, we were able to observe a clear field-effect modulation of $\Delta^{*}$ reflecting its gate-voltage dependence on the Fermi velocity in the 2DHG.  This experiment was performed with QPCs hundreds of nm away from the S/Sm interface. Besides allowing for an electrostatic control over the carrier density in the proximitized 2DHG, such a large distance prevents direct tunneling into the superconducting contact, thereby avoiding the possibility of overestimating the size and hardness of the induced-gap. In conclusion, these findings offer key insights into the proximity effect in hole-based semiconductors, aligning with and supporting recent theoretical developments in the field \cite{Babkin2025, Pino2025}.

\begin{acknowledgements}
This work has been supported by the PEPR ROBUSTSUPERQ (Grant No. ANR-22-PETQ-0003). We would like to thank Axel Leblanc and Sergey Frolov for useful discussions.
\end{acknowledgements}

\subsection*{Author contributions}
Device design and fabrication were performed by E.K. with input from S.D.F. and help from G.T., and C.T. who developed some of the essential fabrication processes. E.K. carried out all of the experiments with help from C.T., B.B., S.Z., and V.S. The results were interpreted and analyzed by E.K., S.D.F., and M.H, with input from B.B., S.Z., V.S., R.M., and F.L. J.-M.H. grew the heterostructure. E.K. and S.D.F. wrote the manuscript with input from all co-authors. The project was supervised by S.D.F.

\subsection*{Conflict of interest statement}
The authors declare no competing financial interest.

\subsection*{Data availability}
All the data and analysis are available at: .............

\setcounter{figure}{0}
\renewcommand{\thefigure}{S\arabic{figure}} \setcounter{figure}{0}

\appendix

\section{DEVICE DETAILS} 
% device details
Devices are electrically isolated from each-other by etching away the conductive Ge layer using reactive ion etching (RIE) of CF$_{4}$ + Ar to create individual mesas typically designed to be \SI{3}{\micro\meter} wide by \SI{16}{\micro\meter} long. The S/Sm interface is made on either side of the mesa by etching the upper SiGe barrier layer using the same RIE etch then transferring the sample to a metal deposition chamber briefly exposing it to ambient air. Ar etching is performed before Al deposition of \SI{50}{\nano\meter} directly over the quantum well. For the chip containing D2 and D3, an O$_2$ plasma surface treatment is made with an inductively coupled plasma power \SI{1000}{\watt}. Then \SI{15}{\nano\meter} of alumina oxide is grown by ALD at \SI{280}{\degree}C to separate the active device layers from the first gate layer of Ti/Au (\SI{3}{\nano\meter}/\SI{62}{\nano\meter}). A second gate layer is deposited after another oxide layer of the same thickness to form the accumulation gates which cover all the regions not covered by the first gate layer.

\section{HALL BAR MEASUREMENTS} 

Hall bar devices are fabricated from the same wafers as used for D1 and D2 with a similar fabrication procedure \cite{KiyookaPhD2025}. These devices were measured at \SI{4}{\kelvin} in four-point resistance measurements in a perpendicular magnetic field yielding a measured current, transverse and longitudinal voltage drop which is then converted to a carrier density ($\rho_{_{\mathrm{2D}}}$) and mobility ($\mu$) shown in Fig. S1. From this data, we calculate a square resistance of $R_{\square} \approx $ \SI{60}{\ohm} and $R_{\square} \approx $\SI{160}{\ohm} for devices D1 and D2 in the strong accumulation regime, respectively. 

\begin{figure}[h!]
	\centering
	 \includegraphics[width=\linewidth]{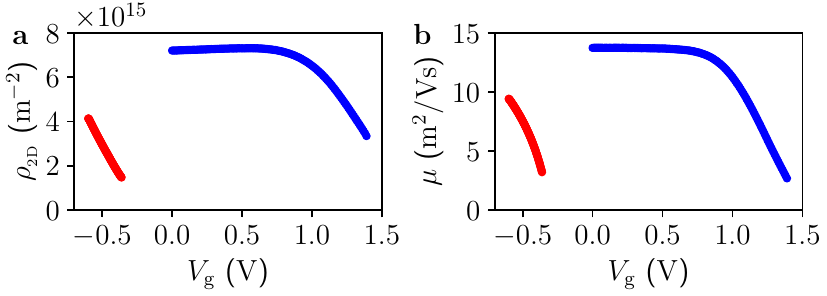}
\caption{Hall bar measurement of (a) the carrier density ($\rho_{_{\mathrm{2D}}}$) and (b) the mobility ($\mu$) as a function of gate voltage for a Hall bar device similar to D1 that has not undergone an oxygen treatment step (blue) and for D2 that has undergone oxygen treatment (red).}
\end{figure}

\section{SERIES RESISTANCE EFFECTS}

In the two-point measurement configuration of our experiment, the presence of a sizable series resistance has two to be taken into account by rescaling the applied bias voltage and the  conductance. 

For the rescaling of the bias voltage we use the relation
$V_{\mathrm{sd}} = V_{\mathrm{DC}} -  I_{\mathrm{DC}} R_{\mathrm{series}}$,
where $V_{\mathrm{sd}}$ is  the actual voltage drop over the device,
$V_{\mathrm{DC}}$ the externally applied DC bias voltage (at the top of the fridge), 
$I_{\mathrm{DC}}$ the DC current through the circuit, and $R_{\mathrm{series}}$ the series resistance  (including the measurement wiring with its low-pass filters, the input  resistance of the current-voltage converter, and a series resistance contribution coming from the device ohmic contacts and the 2DHG). 
Then for the rescaling of the differential  conductance, we use $dI/dV_{\mathrm{sd}} = 
1 / (dV/dI -  R_{\mathrm{series}})$, where $dV/dI$ is the differential conductance measured by lock-in technique, i.e. $dV/dI =V_{AC}/I_{AC}$, where $V_{AC}$ is the externally applied voltage modulation and $I_{AC}$ the measured current modulation. Hence, the linear conductance $G$ corresponds to $dI/dV_{\mathrm{sd}}$ measured at $V_{DC} = 0$. 

\section{SERIES RESISTANCE ESTIMATION FOR DEVICE D1}

\begin{figure}[h!]
	\centering
	 \includegraphics[width=\linewidth]{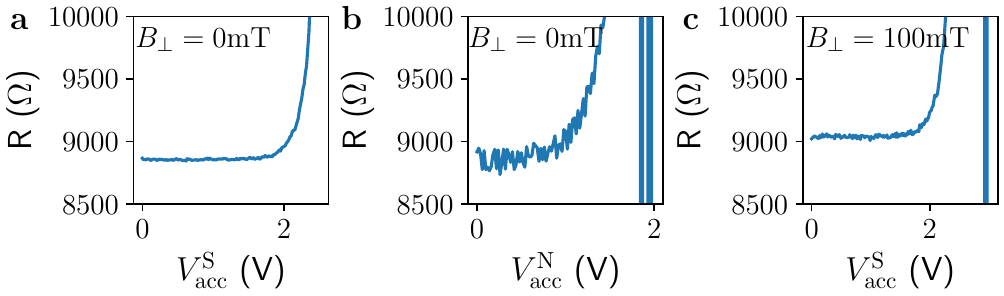}
\caption{Device D1 two-point resistance measured as a function of gate voltage  $V_{\mathrm{acc}}^{\mathrm{S}}$ or  $V_{\mathrm{acc}}^{\mathrm{N}}$) with zero voltage applied to the other gates. In the strong accumulation regime the resistance saturates at \SI{8.85}{\kilo\ohm} for zero applied magnetic field (a,b) and \SI{9.04}{\kilo\ohm} for a \SI{100}{\milli\tesla} perpendicular magnetic field (c).} 
\end{figure}

We set the series resistance equal to the two-point resistance value measured at full accumulation as shown in Fig. S2. We find $R_{\mathrm{series}} = $ \SI{8.85}{\kilo\ohm} and $R_{\mathrm{series}} = $ \SI{9.04}{\kilo\ohm} for $B_{\perp} = $\SI{0}{\milli\tesla} and $B_{\perp} = $\SI{100}{\milli\tesla}, respectively. This resistance corresponds to the sum of the resistance associated with the measurement circuit (low-temperature RC filters, wiring, and input resistance of I-V converter, totally around \SI{7.7}{\kilo\ohm}) and a remaining contribution coming from the device itself (contact resistances and 2DHG). Given that the mesa is $\sim$\SI{16}{\micro\meter} long by $\sim$\SI{2.4}{\micro\meter} wide with a square resistance of $R_{\square}\sim$ \SI{60}{\ohm} at full accumulation, we estimate the contribution from the 2DHG to be $\sim$\SI{0.4}{k\ohm}. Under an applied out-of-plane magnetic field the aluminum leads become resistive causing the observed \SI{190}{\ohm} increase of series resistance. 

\section{CONDUCTANCE QUANTIZATION FITTING}

\begin{figure}[h!]
	\centering
	 \includegraphics[width=\linewidth]{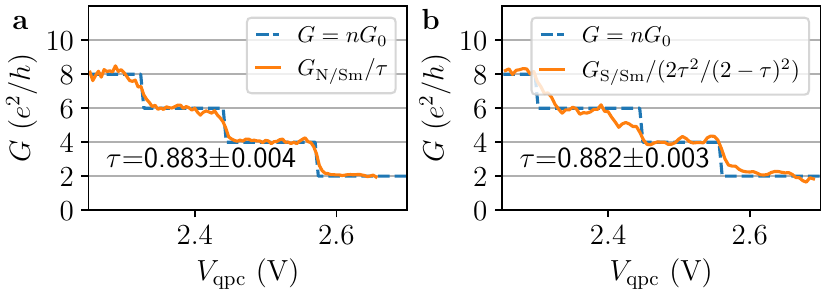}
\caption{(a) Measured normal-state conductance (orange curve) rescaled by a transmission coefficient $\tau$ to align with a conductance staircase with plateaus at integer multiples of $G_0$ (dashed line). (b) Measured Andreev-enhanced conductance (orange curve) rescaled by the expected conductance enhancement factor ($2\tau^2/(2-\tau)^2$) to align with a conductance staircase with plateaus at integer multiples of $G_0$ (dashed line).}
\end{figure}

Figure S3 shows the first four plateaus of the normal (a) and Andreev-enhanced conductance (b) normalized by $\tau$ and by $2\tau^2/(2-\tau)^2$, respectively (orange solid lines). The fit parameter $\tau$ is adjusted to minimize the deviation from a conductance staircase with abrupt $G_0$ steps (dashed lines). The abrupt steps align with the maxima in the first derivative of the corresponding data set. Both fits yield $\tau = 0.88$. 

\section{FINITE-BIAS COMPARISON OF THE CONDUCTANCE ENHANCEMENT}

Figure S5 reproduces the averaged data from Fig. 1(f) where the $dI/dV_{\mathrm{sd}}$ trace from the second conductance plateau has been divided by a factor two. Following this rescaling, the two $dI/dV_{\mathrm{sd}}$ traces fall approximately on each other. This shows that the first two modes provide basically the same contribution to the Andreev-enhanced differential conductance up to bias voltages well above $\Delta^{\mathrm{Al}}_{\mathrm{BCS}}/e$, consistent with an equal and bias-independent transmission coefficient. 

\begin{figure}[h!]
	\centering
	 \includegraphics[width=\linewidth]{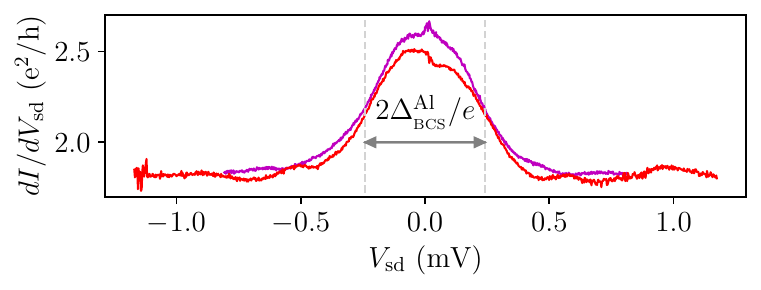}
\caption{Data from Fig. 1(f) with the $dI/dV_{\mathrm{sd}}$ from the second conductance plateau divided 2.}
\end{figure}

\section{FULL IMAGE OF DEVICE D2}

Figure S6 shows a full SEM image of the D2 device. On the left side, we notice the first Al contact with two facing QPCs. The area delimited by a red rectangle corresponds to the device portion shown in Fig. 2(a). On the right side, we distinguish two additional QPCs facing the second Al contact. The central C gate runs horizontally from the first to the second Al contact. A sufficiently positive gate voltage applied to this gate, defines two parallel channels. The top (bottom) channel, which corresponds to right (left) channel in Fig. 2(a), is nominally 1.15 (0.74) \SI{}{\micro\meter} wide and \SI{17}{\micro\meter} long. 

\begin{figure}[h!]
	\centering
	 \includegraphics[width=\linewidth]{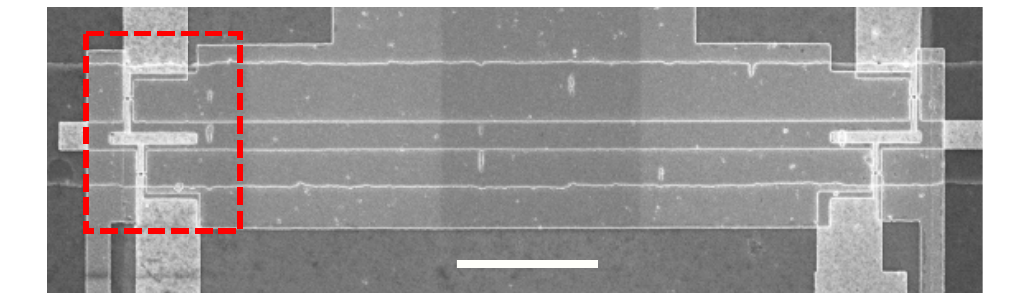}
\caption{SEM image of the D2 device (rotated by 90$^{\circ}$ relative to Fig. 2 (a)) showing both aluminum contacts. The red box indicates the device region shown in Fig. 2(a) with the investigated QPCs. Scale bar: \SI{3}{\micro\meter}. }
\end{figure}

\section{SERIES RESISTANCE ESTIMATION FOR DEVICE D2}

\begin{figure}[h!]
\centering
\includegraphics[width=\linewidth]{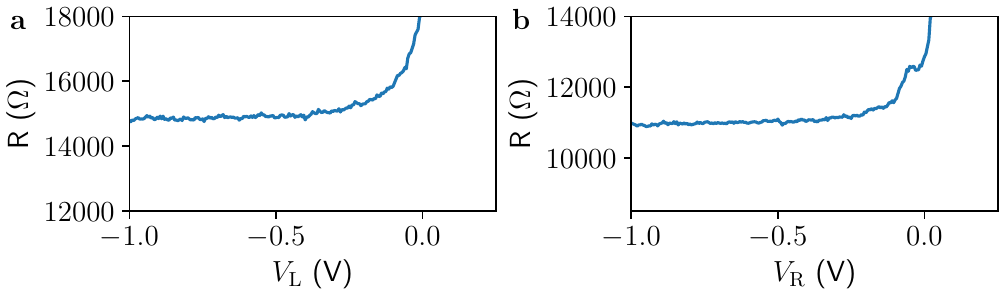}
\caption{Two-point resistance measurement as a function of the QPC gates ($V_{\mathrm{L,R}}$) in device D2 with $V_{\mathrm{C}} = $ \SI{0.25}{\volt}, $V_{\mathrm{acc}}^{\mathrm{S,N}} = $ \SI{-2}{\volt}, i.e. in the strong-accumulation regime. In each case, the other QPC is completely pinched off. With the QPC fully open the resistance saturates  at  (a) $R^{\mathrm{L}}_{\mathrm{series}} = $ \SI{14.85}{\kilo\ohm} and (b) $R^{\mathrm{R}}_{\mathrm{series}} = $ \SI{10.94}{\kilo\ohm}.}
\end{figure}

As shown in Fig. S6, D2 has a significantly higher resistance than D1 having for the left side $R_{\mathrm{series}}^{\mathrm{L}} = $ \SI{15}{\kilo\ohm} and for the right side $R_{\mathrm{series}}^{\mathrm{R}} = $ \SI{11}{\kilo\ohm}. These higher values are believed to be due to the higher square resistance ($R_{\square}\sim$\SI{160}{\ohm}), and a narrower device geometry. The mesa is $\sim$\SI{17}{\micro\meter} long, but now being cut by the central gate the width is only $w_{\mathrm{L}}\sim$ \SI{.74}{\micro\meter}  or $w_{\mathrm{R}}\sim$ \SI{1.15}{\micro\meter} wide (see Fig. S5). These values imply devices resistances of $R_{\mathrm{dev}}^{\mathrm{R}} = $ \SI{3.6}{\kilo\ohm} and $R_{\mathrm{dev}}^{\mathrm{L}} = $ \SI{2.3}{\kilo\ohm}, which is not enough to add up to the measured value (RC filters again total to \SI{7.7}{\kilo\ohm}). However, we have been assuming the entire mesa is conducting up to the very edge. For $w_{\mathrm{L}}\sim$\SI{.74}{\micro\meter}, if some portion of the edge is not conducting or damaged it will significantly increase the device resistance which appears to be the case from the rough edges and local defects in the mesa of Fig. S5. %Additionally, since we change the accumulation gate voltage $V_{\mathrm{acc}}^{\mathrm{S}}$, the series resistance was measured for each side and at each value of $V_{\mathrm{acc}}^{\mathrm{S}}$ used in the main text Fig. 3.

\section{TUNNELING SPECTROSCOPY} 

\begin{figure}[h!]
	\centering
	 \includegraphics[width=\linewidth]{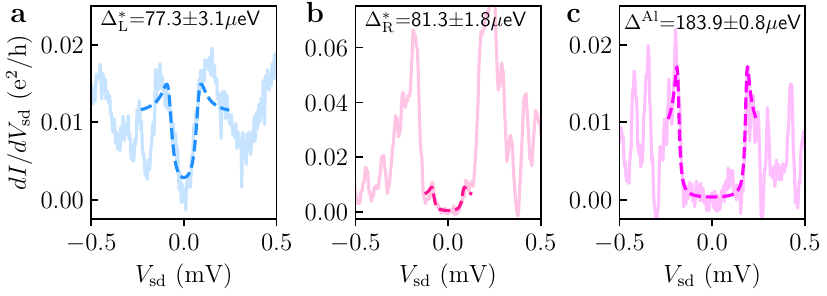}
\caption{Tunneling spectroscopy data sets (solid lines) and corresponding fits to a broadened BCS DOS function (dashed lines), illustrating the methodology to extract the superconducting gaps $\Delta^{*}_{\mathrm{L,R}}$ and $\Delta^{\mathrm{Al}}$ (Fitting voltage ranges:  $[\pm$\SI{120}{\micro\eV}$]$ and  $[\pm$\SI{240}{\micro\eV}$]$, respectively). Panel (a) shows data from the left QPC, while panels (b) and (c)  show data from the right QPC. The given  uncertainties in the extracted gaps are simply fit uncertainties.}
\end{figure}

% Dynes fitting details 
Gap features in the tunneling spectroscopy measurements are fit to a broadened BCS DOS function of the form 
\begin{equation*} 
(dI/dV_{\mathrm{sd}}) = N_{o}|\Re{[(E-i\Gamma)/\sqrt{(E-i\Gamma)^2-\Delta^2}]}|
\end{equation*}
where $E=eV_{sd}$, $\Delta$ is the superconducting gap to be fit,  $\Gamma$ is a phenomenological fit parameter (usually known as Dynes parameter), and $N_{o}$ is a fit proportionality constant. As shown in Fig. S7, the two gap features $\Delta^{*}_{\mathrm{L,R}}$ and $\Delta^{\mathrm{Al}}$ are extracted by fitting the tunneling spectroscopy data varying voltage ranges.

The data in Fig. 3 is extracted from 2D  plots of the differential-conductance as a function of $V_{\mathrm{L}}$ (or $V_{\mathrm{R}}$ and $V_{\mathrm{sd}}$ where we fit  $dI/dV_{\mathrm{sd}}$ traces that have a normal-state resistance $R_{n}$ ranging from 0.5 to 2.6 \SI{}{\mega\ohm}. $R_{n}$ is taken as the inverse of the averaged high-bias ($V_{\mathrm{sd}} > \Delta^{\mathrm{Al}}_{_\mathrm{BCS}}/e$) differential conductance.  The full datasets are shown in Fig. S10-S11.  For each value of $V_{\mathrm{acc}}^{\mathrm{S}}$, the ensemble of gap values obtained from fitting the $dI/dV_{\mathrm{sd}}$ traces that fulfill the above condition then averaged yielding a single data point of Fig. 3 with an uncertainty corresponding to the standard deviation. Our fits give different values of the Dynes parameter with  $1\sigma$ distributions $\Gamma_{\mathrm{L}} \in$ [3,43] \SI{}{\micro\eV}, $\Gamma_{\mathrm{R}} \in$ [7,10] \SI{}{\micro\eV}, and $\Gamma_{\mathrm{Al}} \in$ [13,17] \SI{}{\micro\eV}.

\section{INDUCED GAP OF DEVICE D1}

\begin{figure}[h!]
	\centering
	 \includegraphics[width=\linewidth]{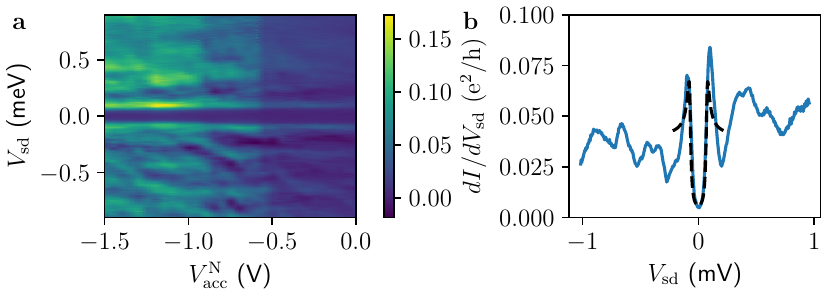}
\caption{Measurement of the induced gap in device D1 using the voltage on the normal-side accumulation gate on the normal side, $V_{\mathrm{acc}}^{\mathrm{N}}$, as varying parameter to average out mesoscopic fluctuations $V_{\mathrm{acc}}^{\mathrm{S}} = $  \SI{-1.5}{\volt}, $V_{\mathrm{qpc}}^{\mathrm{L}} = $  \SI{2.937}{\volt}, $V_{\mathrm{qpc}}^{\mathrm{R}} = $  \SI{2.706}{\volt}. (a) Conductance through the QPC as a function of DC bias ($V_{\mathrm{sd}}$) and the normal accumulation gate ($V_{\mathrm{acc}}^{\mathrm{N}}$). (b) Line-cut average in bias over the entire $V_{\mathrm{acc}}^{\mathrm{N}}$ range of (a) with the fit to the Dynes function giving $\Delta^{*}_{\mathrm{dev1}} = $ \SI{73}{\micro\eV}. }
\end{figure}

In Fig. S8, the induced gap is measured for D1 in the low conductance regime $G\ll G_{0}$. This separate measurement was necessary since the $V_{\mathrm{sd}}$ line-cut traces of the main text Fig. 1 do not have sufficiently high resolution to be used for fitting to a broadened BCS DOS function. This conductance dataset with independent variables $V_{\mathrm{sd}}$ and  $V_{\mathrm{acc}}^{\mathrm{N}}$ is measured at particular values of $V_{\mathrm{qpc}}^{\mathrm{L,R}}$ as this was the region where the most symmetric BCS peaks were identified. Averaging over $V_{\mathrm{acc}}^{\mathrm{N}}$ allowed us to average over mesoscopic fluctuations to yield a trace more easily fit by the Dynes function yielding $\Delta^{*}_{\mathrm{dev1}} = $ \SI{73}{\micro\eV} comparable to the extracted values of D2.  
\section{STABILITY DIAGRAMS USED TO EXTRACT THE GATE VOLTAGE DEPENDENCE OF THE INDUCED SUPERCONDUCTING GAP}

\begin{figure}[h!]
	\centering
	 \includegraphics[width=\linewidth]{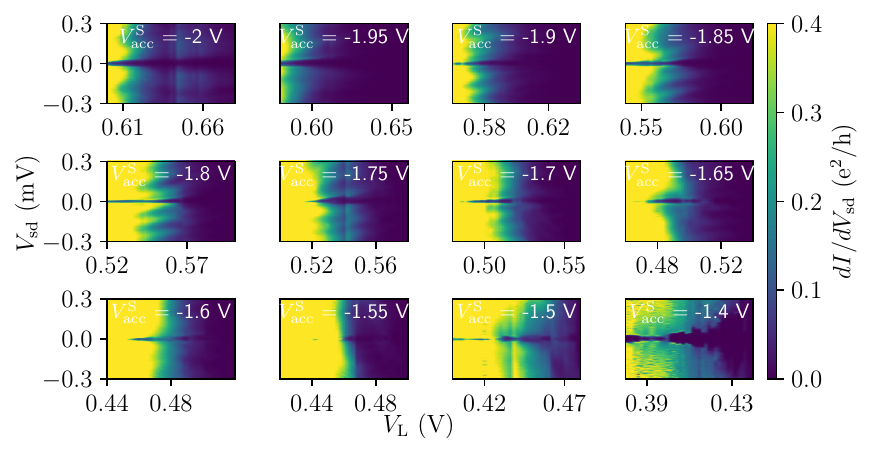}
\caption{Differential conductance, $dI/dV_{\mathrm{sd}}$, for the left QPC on device D2 as a function of  $V_{\mathrm{L}}$ and $V_{\mathrm{sd}}$ for different values of  $V^{\mathrm{S}}_{\mathrm{acc}}$ with $V_{\mathrm{C}} = $ \SI{0.15}{\volt} and $V_{\mathrm{R}} = $\SI{1}{\volt}. }
\end{figure}

\begin{figure}[h!]
	\centering
	 \includegraphics[width=\linewidth]{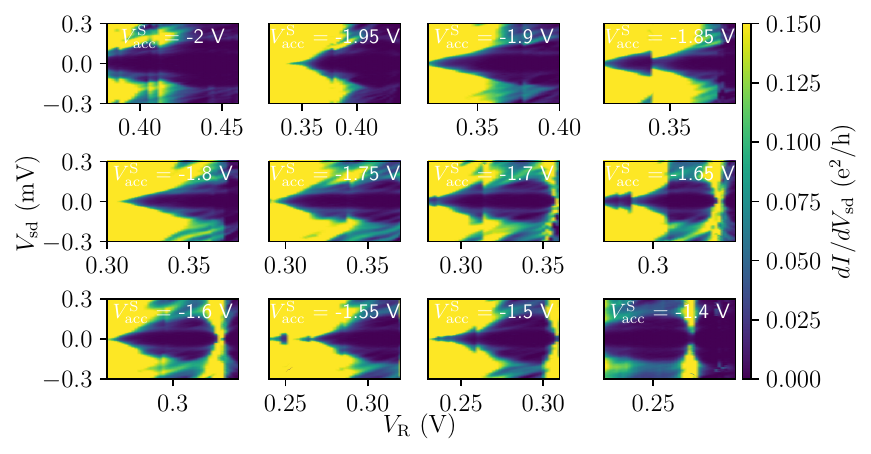}
\caption{Differential conductance, $dI/dV_{\mathrm{sd}}$, for the right QPC on device D2 as a function of $V_{\mathrm{R}}$ and  $V_{\mathrm{sd}}$ for different values of  $V^{\mathrm{S}}_{\mathrm{acc}}$ with $V_{\mathrm{C}} = $ \SI{0.35}{\volt} and $V_{\mathrm{L}} = $\SI{1}{\volt}.}
\end{figure}

The data in Fig. 3 are extracted from individual 2D plots of $dI/dV_{\mathrm{sd}}$ as a function of QPC gate voltage ($V_{\mathrm{L}}$, $V_{\mathrm{R}}$, and bias voltage ($V_{\mathrm{sd}}$), each taken with different values of $V^{\mathrm{S}}_{\mathrm{acc}}$. These datasets are plotted in Fig. S9 (D2, left QPC), and Fig. S10 (D2, right QPC). 

The reduction in $\Delta^{*}_{\mathrm{L}}$ as a function of $V^{\mathrm{S}}_{\mathrm{acc}}$ can be visualized directly in Fig. S9 by a progressive shrinking of the horizontal blue region of suppressed $dI/dV_{\mathrm{sd}}$ centered around $V_{\mathrm{sd}} = $ \SI{0}{\volt}. In the last datasets for  $V^{\mathrm{S}}_{\mathrm{acc}} = $ \SI{-1.5}{\volt} and \SI{-1.4}{\volt}, the device appears to enter into a regime characterized by the emergence of quantum-dot physics. For this reason, datasets for $V^{\mathrm{S}}_{\mathrm{acc}} > $ \SI{-1.55}{\volt} were not used in the study of  $\Delta^{*}_{\mathrm{L}}(V^{\mathrm{S}}_{\mathrm{acc}})$. Figure S10 shows both the parent superconducting gap and the nested induced gap, with no dependence on $V^{\mathrm{S}}_{\mathrm{acc}}$. In addition, some of the datasets show superimposed features most likely originating from gate-dependent Andreev-bound states localized in the QPC region. These cause an apparent collapsing of the gaps and concomitant peak in $dI/dV_{\mathrm{sd}}$ at certain values of $V_{\mathrm{R}}$ where the Andreev bound states cross the Fermi energy. 

\section{DEVICE D3 MEASUREMENTS}

\begin{figure}[hb!]
	\centering
	 \includegraphics[width=\linewidth]{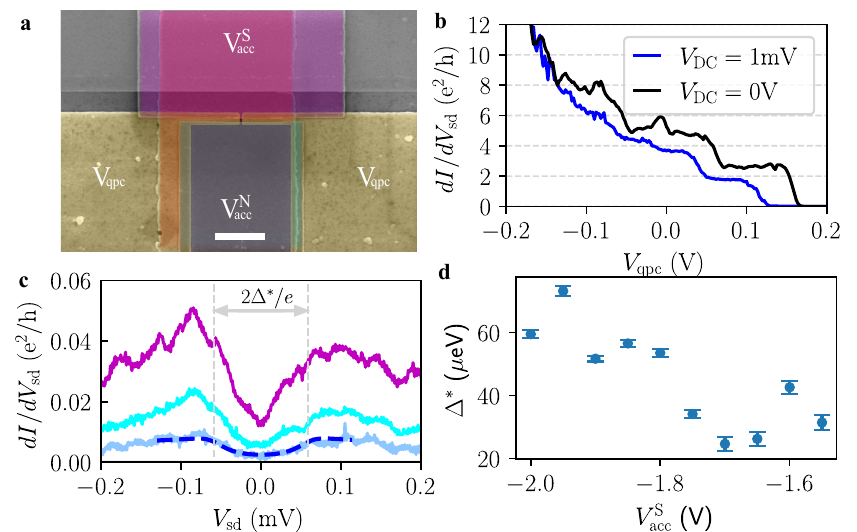}
\caption{Measurement results from device D3 ($d\sim$ \SI{400}{\nano\meter}). (a) False-color SEM image with similar coloring as in Fig. 1(a) and Fig. 2(a) (scale bar: \SI{1}{\micro\meter}). (b) Differential conductance, $dI/dV_{\mathrm{sd}}$, as a function of QPC gate voltage, $V_{\mathrm{qpc}}$, for DC bias $V_{\mathrm{DC}} = 0$ and 1 \SI{}{\milli\volt}. (c) Tunneling spectroscopy measurements with a fit (dashed line) to extract the induced superconducting gap $\Delta^{*} \approx $ \SI{60}{\micro\eV}. (d) Induced gap as a function of the accumulation gate voltage $V_{\mathrm{acc}}^{\mathrm{S}}$.}
\end{figure}

In Fig. S11, we show a summary of measurements on a third device D3 made on the same chip containing D2. Device D3 consists of one split-gate QPC at a distance $d\sim$ \SI{400}{\nano\meter} from the nearby superconducting contact, as shown in Fig. S11(a). Consistently with previous observations, D3 exhibits Andreev-enhanced conductance  ($dI/dV_{\mathrm{sd}}$ at $V_{\mathrm{DC}} = $ \SI{0}{\milli\volt}) above the normal-state conductance ($dI/dV_{\mathrm{sd}}$ at $V_{\mathrm{DC}} = $ \SI{1}{\milli\volt}) by a similar factor $\sim 1.4$, with estimated $R_{\mathrm{series}} \approx $ \SI{9}{\kilo\ohm} (see Fig. S11(b)). 

Similarly to Fig. 3, we fit tunneling spectroscopy data to a broadened BCS DOS function for different values of the QPC gate voltage $V_{\mathrm{qpc}}$ and the accumulation gate voltage $V_{\mathrm{acc}}^{\mathrm{S}}$ as detailed in Appendix I. The largest value of the fitted induced gap is $\Delta^{*} \sim $ \SI{70}{\micro\eV} decaying by at least a factor of three with increasing gate voltage. These results are overall consistent with those shown in Fig. 3 from D2. 

\bibliographystyle{apsrev4-2}
\bibliography{refs}

\end{document}